\def\year{2022}\relax
\documentclass[letterpaper]{article} % DO NOT CHANGE THIS

\usepackage[english]{babel}
\usepackage{amsmath} %amssymb
\usepackage{bbold}
\usepackage[algo2e]{algorithm2e}
\usepackage{xcolor}
\usepackage[normalem]{ulem}
\usepackage{bbm}

\newcommand{\ac}{a}
\newcommand{\co}{c}
\newcommand{\Co}{\mathcal{C}}
\newcommand{\Ac}{\mathcal{A}}

\DeclareRobustCommand{\Paragraph   }[1]{\noindent \textbf{#1}}

\usepackage{aaai22}  % DO NOT CHANGE THIS
\usepackage{times}  % DO NOT CHANGE THIS
\usepackage{helvet}  % DO NOT CHANGE THIS
\usepackage{courier}  % DO NOT CHANGE THIS
\usepackage[hyphens]{url}  % DO NOT CHANGE THIS
\usepackage{graphicx} % DO NOT CHANGE THIS
\usepackage{natbib}  % DO NOT CHANGE THIS AND DO NOT ADD ANY OPTIONS TO IT
\usepackage{caption} % DO NOT CHANGE THIS AND DO NOT ADD ANY OPTIONS TO IT
\DeclareCaptionStyle{ruled}{labelfont=normalfont,labelsep=colon,strut=off} % DO NOT CHANGE THIS
\usepackage{algorithm}
\usepackage{algorithmic}

\usepackage{booktabs}

\usepackage{newfloat}
\usepackage{listings}
\floatstyle{ruled}
\newfloat{listing}{tb}{lst}{}
\floatname{listing}{Listing}
\usepackage{subcaption}

\definecolor{mygreen}{RGB}{80, 220, 100}

\definecolor{mygmagenta}{RGB}{181, 45, 116}

\definecolor{myyellow}{RGB}{204, 204, 0}

\title{Teaching Drones on the Fly: Can Emotional Feedback Serve as Learning Signal for Training Artificial Agents?}
\author {
    Manuela Pollak,
    Andrea Salfinger,
    Karin Anna Hummel
}
\affiliations {
    Institute for Telecooperation\\
    Johannes Kepler University Linz\\
    Altenberger Strasse 69\\
    4040 Linz, Austria\\
    manuela.pollak@jku.at, andrea.salfinger@cis.jku.at, karin\_anna.hummel@jku.at
}
\usepackage{bibentry}
\begin{document}

\maketitle
\thispagestyle{plain}

\begin{abstract}
We investigate whether \emph{naturalistic emotional human feedback} can be directly exploited as a \emph{reward signal} for training artificial agents via \emph{interactive human-in-the-loop reinforcement learning}. To answer this question, we devise an experimental setting inspired by animal training, in which human test subjects interactively teach an emulated drone agent their desired command-action-mapping by providing emotional feedback on the drone's action selections. We present a first empirical proof-of-concept study and analysis confirming that human facial emotion expression can be directly exploited as reward signal in such interactive learning settings. Thereby, we contribute empirical findings towards more naturalistic and intuitive forms of %machine teaching 
%\karin{reward-based machine learning}
%\andrea{reinforcement learning}
reinforcement learning especially designed for non-expert users.
\end{abstract}

\section{Introduction}

We study the research question \emph{can emotional feedback serve as learning signal for intuitively training artificial agents} on the use case of human-drone interaction. We present an experimental setting in which a user intuitively teaches a drone the meaning of commands such as gestures that can be detected %recognized
by vision-based recognition~\cite{MinhDang.2020, Hummel.2019}. Learning is based on emotional feedback, a \emph{naturalistic mode of interaction} %(such as facial expression, verbal responses etc.) that 
humans are familiar with from conventional human-to-human and human-to-animal interactions. 

\Paragraph{Learning Approach.} We establish an analogy to \emph{animal training}, where animals are instructed to learn the correct action for a given command. Here, learning is based on evaluative feedback from the animal's instructor, traditionally in form of tangible rewards such as the animal's preferred food~\cite{VieiradeCastro.2021}, 
but often also in form of the instructor's emotional response, especially for social animals such as canines. Over time, the animal learns to correlate the received reward with the correct action. Similarly, an artificial agent -- here, a drone -- should interpret the emotional feedback of the user to learn. 

The ability to process emotional cues can be considered as a key requirement for any socially intelligent agent. At the same time, the intuitiveness of emotional feedback allows to include users with little technical background in the training/rewarding task. Thus, we believe that providing a method to interpret affective evaluative human feedback represents a cornerstone in social human to machine interaction.

\begin{figure}[!t]
	\centering
	\includegraphics[width=\columnwidth]{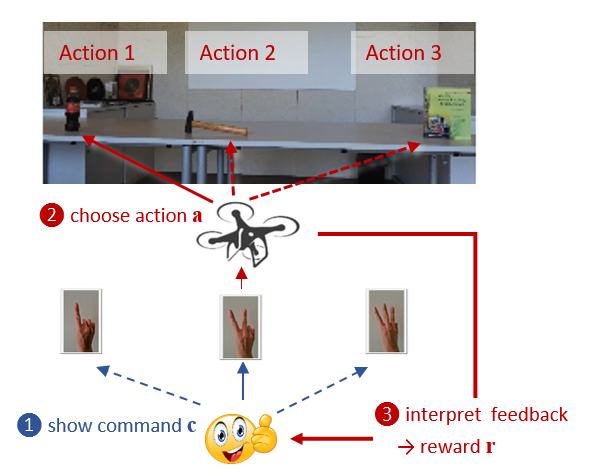}
	\caption{The \emph{interactive machine learning loop} of our drone training experiment: The user issues a gesture command $c$ upon which the drone chooses an action $a$. The user provides emotional feedback which the drone interprets to derive a reward. %This sequence is repeated until the user is satisfied. 
	In this contribution, we specifically focus on deriving rewards from interpreting naturalistic emotional feedback.}
	\label{fig:webappexp}
\end{figure}

\Paragraph{Contributions.} %Our contributions are: 
(i) We present the design and implementation of a real-time interactive machine learning (IML) system based on reinforcement learning (RL), where rewarding is provided by human emotional feedback. We envision an artificial agent that makes use of a camera and facial emotion recognition and exemplify the agent by a small indoor drone that requires to learn a command-action mapping. Among the core features of the system, a \emph{rewarding algorithm} is developed that converges fast and allows to cope with different expressiveness of individual facial emotions. (ii) To study whether emotional feedback can provide 
a useful learning signal for the drone, we conduct an experimental user study, where we mimic the drone's actions upon receiving a command as sketched in Fig.~\ref{fig:webappexp}. Our first results indicate that the approach is feasible. %to learn the correlation of command and action by emotional feedback.

\section{Background and Related Work}

We first motivate why emotions play an important role in human-robot interaction (HRI), summarize major results on the role of feedback in agent learning, and give an overview of emotion recognition as feedback signal and direct, naturalistic form of interaction.

%Here, we may start to list and summarize related work on the topic ... 

\Paragraph{The Role of Emotions in HRI.}  
A large fraction of research studies on the role of emotions in HRI settings has focused on %\emph{human factors} of the interaction by 
investigating the human's preferences of interaction modalities to better understand the robot's behavior, such as the studies described in~\cite{Reyes.2019,Thunberg.2021}. Another field of study examines how typically humanoid robots' emulated emotion expressions affect human performance. For example, studies on robot companions for children investigated the effect of the robot's displayed emotional response as a \emph{feedback and motivation signal} for children's learning performance~\cite{Ahmad.2019}. 

Yet, also the artificial agents' ability to correctly interpret and react to human emotions has been acknowledged as a key requirement for socially intelligent agents and human-machine interaction in the fields of cognitive robotics, artificial intelligence (AI), and HRI, such as described in~\cite{Chakraborti.15.07.2017,Biundo.2016,Hornle.2017}. The ability of a robot to infer the emotional state of a human reliably is a desired goal. Research has targeted the creation of emotional models and classification schemes of human emotions, and emotion recognition such as facial emotion recognition (FER) and analysis of body posture as well as brain activity as surveyed in~\cite{er2020}, yet only few works have focused on the feasibility of using emotion as reliable signal for feedback-based machine learning.

%In the present work, we are essentially interested in studying the opposite direction of the interaction: How can the \emph{machine} employ the human's emotional feedback as learning signal? \todo{is there no work on using emotion to signal something to a robot?}

\Paragraph{Feedback in Agent Learning.} In his characterization of the twelve potential roles that emotion could play in AI, \cite{Scheutz.2004} pointed out the possibility of \emph{using emotion for learning}, for example by using emotional evaluations as $Q$-values in reinforcement learning~(RL), as we will attempt in this work. RL represents the machine learning paradigm employing learning from \emph{evaluative feedback}.\footnote{\emph{Evaluative} feedback indicates how good/bad the taken decision was. This is opposed to \emph{imperative} feedback that states what the \emph{correct} decision would have been (i.e., the paradigm of \emph{supervised} machine learning), or \emph{descriptive} feedback, which states \emph{how} the learner should adapt its behavior~\cite{Sumers.2021}.} Classically, RL agents compute a reward score from their interactions with the environment, using a pre-defined reward function  to provide the feedback signal for the agent. This way, the desired behavior is reinforced by rewarding. However, this reward function is specific for the given task, and
consequently the agent’s learning capability often hinges on the
designer’s capability of engineering an appropriate reward function for the problem at hand~\cite{PedroSequeira.2013,Russell.2019b}. 
%Yet, \andrea{designing an appropriate reward function} as they may miss important goals~\cite{PedroSequeira.2013}. %\remark{for me a bit coarse now, would still make it more specific that we're talking about the problem of \emph{engineering} the \emph{reward function} here...}. 
To allow for more generic open-world learning, approaches such as presented in~\cite{Thomaz.2008,Knox.2009,MacGlashan.2017} %already investigated how humans could directly train the agent and guide its learning, by monitoring the agent and providing feedback on its behavior, termed
propose learning through \emph{human reinforcement}, i.e., enable a human tutor to assess the agent's behavior and to provide direct feedback. In past works, this feedback has to be provided in the form of predefined, numeric reward scores, which is not the natural interaction style we are aiming for. %heading for. 

\Paragraph{Naturalistic Human Feedback.} %\andrea{\sout{In human-machine-teams, collaboration} 
Human-robot interaction (HRI) can reach a next level if an artificial agent is capable of interpreting the human's feedback in a more naturalistic manner. In terms of the RL paradigm, the agent should be capable of inferring its (numeric) reward scores from humans' natural communication channels. %(speech, facial expression etc.).  
%Using text as a communication channel for deriving rewards \remark{hm..., at this point not entirely sure anymore whether we can put it like that since they actually learned to infer the reward \emph{function} in this IRL setting},
For example, \cite{Sumers.2021} devised a learning approach using \emph{inverse RL} %to learn humans' latent reward function from naturalistic 
to learn humans' latent reward function from (unconstrained, thus naturalistic)  textual feedback captured during a human-human collaborative game play setting. Other research work studied the utilization of emotional evaluations as rewards, as \cite{Scheutz.2004} proposed. \cite{Broekens.2007} indeed already experimented with incorporating humans' emotional response as a reward signal in RL.  %estimating the human tutor's reward by interpreting their recorded emotional expression.
However, in this approach emotional feedback is used only as an \emph{additional reward} added on top of a base reward signal the agent obtained from its virtual environment (thus, essentially only makes the base reward signal more accurate). This is due to the chosen RL algorithm based on neural networks, which requires a large number of training iterations and hence is not amenable to continuous human feedback. 

%Differently, in the present work we seek to evaluate whether the human's emotional response could be utilized as the \emph{sole source of feedback} on the %robot's 
%artificial agent's performance. %(which would represent a generic reward function for any types of open-world learning scenarios), 
\Paragraph{Our approach} is similar in spirit to recent endeavors on learning rewards from \emph{unconstrained naturalistic human feedback} such as~\cite{Sumers.2021}. Yet, we extend the aforementioned related work by utilizing the human's emotional response as the \emph{sole source of feedback} on the artificial agent's performance. To overcome the problem of long training efforts, we propose to choose an RL approach well suitable for real-time learning, notably multi-armed bandits that require only a few %fewer 
training iterations until convergence. 
%This approach is expected to }
%that might 
%be more suitable for \emph{real-time machine learning} %human-machine interaction and training 
%as it requires
%\todo{karin - discuss: use of upper case / lower case in the text ... do we want to use style Multi-Armed Bandits or multi-armed bandits (I am in favor of the latter)}
 %\todo{more on NUI?}
%\Paragraph{Towards Ethically-Sound AI Control.} In a wider sense, our contributions can also be seen in the realm of \emph{ethical AI}, following Russel's line of thought on the ethically sound design of \emph{AI control} algorithms~\cite{Russell.2019b}. Russel advocates that humans ultimately would fail to design the correct reward (i.e., objective) function for any scenario (taking analogy from the classic \emph{King Midas} problem, who wanted that everything he touches turns into gold, without considering the consequences of this objective and thus ultimately starves to death), thus proposing that the reward function should \emph{not} be specified with respect to the agent's current task (as ill-specified reward functions could ultimately lead to harming humans), but the agent's goal should be set on optimizing humans' well-being. As a basic requirement for realizing this objective, artificial agents thus also need to be capable of correctly interpreting humans' emotions (a prerequisite for inferring their well-being), as we will investigate in this work in terms of a dedicated %real-world 
%experiment on human-drone-interaction.

\section{Model and Study Design}

%\karin{COMMENT: I moved longer commented parts to the end of the file (also in other sections/files)}

%\Paragraph{Problem Specification.} 
We study the feasibility of using emotional feedback to train an artificial agent along the use case of \emph{human-drone interaction}, where the drone is the artificial agent as depicted in Fig.~\ref{fig:webappexp}. The user issues a search command by a gesture, which results in the drone navigating to the specific object the user searches for. The drone will learn the correct command-action mapping by reinforcement learning based on emotional user feedback. More formally, we can describe the system by an agent-based model.

\Paragraph{Agent-based Model.} We model the human-drone interaction as a multi-agent system consisting of

\begin{itemize}
    \item a user agent $u \in \mathcal{U}$ that can issue a set of commands $\Co$, 
    \item a drone agent $d \in \mathcal{D}$ that can execute a set of actions $\Ac$,
\end{itemize}

where $|\Co| = |\Ac| = k$, and the sets $\mathcal{U}$ and $\mathcal{D}$ represent the sets of users and drones, respectively. During the \emph{training phase} the drone agent $d$ learns which action ${\ac} \in \Ac$ the user agent $u$ desires when issuing a command $\co \in \Co$. The bijective mapping function $f_u(\co) = \ac$ expresses the personalized command-action mapping for a user $u$. 

\Paragraph{Drone Learning by Emotional Rewards.} %When initiating the interaction with a new user, the drone agent has no knowledge on which of its $k = |\Ac|$ possible actions the user wants to be executed based on the issued command $\co$. 
Initially, the drone does not know the command-mapping $f_u(\co) = \ac$ for any command $\co$ and has to learn it by reinforcements from the respective user $u$. The drone selects an action $\ac \in \Ac$ randomly (independently and identically distributed) and awaits feedback in form of emotional feedback from user $u$, which results in a \emph{reward} $r$.   
Eventually, the drone will learn the desired command-action mapping function $f_u(\co) = \ac$ for all commands $\co$ after experiencing a sequence of interactions with user $u$ that can be described as $\langle (\co^0, \ac^0, r^0), (\co^1, \ac^1, r^1), \ldots \rangle$. Yet, the learning outcome is also effected by learning confounders. 

%Yet, leveraging naturalistic human-drone interaction, we rely on sensor-based signals and inference, which are known learning confounders effecting the learning outcome.

\Paragraph{Learning Confounders.} 
Naturalistic user interfaces rely on sensor-based signals and thus require the implementation of user recognition, command recognition, and feedback recognition and interpretation (feedback in form of text, visual, audio, etc.).  Any such recognition involving the drone hardware is prone to classification errors, thus, impacting machine learning performance. In order to reduce this complexity when studying our research question, we design our experimental study to eliminate confounders, i.e., potential error sources that might interfere with the learning problem, here user and gesture recognition errors and interaction with a real hardware-embodied agent. Once we have evidence that emotional feedback is a valid source for teaching, we will extend our study to the hardware drone.

%\footnote{Once we have shown that emotional feedback is a valid source for teaching a drone, we will extend our study and experiment with the hardware drone.}

\Paragraph{Experimental Study Design.} To study the usefulness of naturalistic emotional feedback and the effectiveness of the rewarding algorithm without side-effects caused by confounders related to the hardware, we create a Web-based emulated drone agent. This way,
we abstract user-drone interactions to deterministic, error-free user login (to emulate user recognition), button-clicks (to emulate gesture-based commands) and video replay of recorded real drone flights (to emulate actions performed by the drone). %These abstractions allow us to record and analyze the human users' interactions and emotional responses while omitting influencing side-effects. 
The user trains the drone by providing emotional feedback to the drone based on its action selection. Further, the user provides labels for every emotional feedback in terms ``positive`` and ``negative`` action assessment. Both the desired command-action mappings and the feedback labels serve as \emph{ground truth data}.
%, together with the desired command-action mapping, which serve as \emph{ground truth data}. 

%-------------------------------------------------------
%\section{Implementation: Learning from Human Reinforcement}
\section{RL by Emotional Signals}
\label{sec:rewardcomputation}
%In the present section, we introduce how we implemented the learning from naturalistic feedback.
%For our experiments, we chose to first focus on facial emotion expression as observations of naturalistic feedback for inferring user's latent "satisfaction" state.

Learning by emotional signals is based on the following steps: (i) capturing emotional feedback as an assessment of the drone's action choice by the drone's onboard camera, (ii) recognizing emotions by analyzing the user's facial expression on video frames resulting in an emotion distribution, and (iii) deriving a reward score based on the emotion distribution for drone actions to finally implement the command-action mapping $f_u(\co) = \ac$, personalized for each user $u$.

\Paragraph{Capturing Emotional Feedback.}
%We use humans' facial emotion expressions as naturalistic feedback signal. 
%Regarding the type of naturalistic feedback employed, we chose to first focus on facial emotional expressions:
The user will try to convey to the drone visually how (dis)satisfied s/he was with the drone's action choice for a predefined feedback time interval. The drone records a video stream of this evaluative feedback, i.e., a time series of video frames, intended as raw input for drone learning. Each detected face is analyzed based on facial emotion recognition. The number of frames to be considered is configurable (sampling fewer frames may lead to missed emotions, sampling more frames leads to higher processing effort).

%hence its raw input for learning consists of a time series of video frames, 
%$\langle frame_0, frame_1, \ldots frame_N \rangle$, 
%which it needs process in real-time to enable a seamless continuous interaction. We next introduce our devised approach for deriving the agent's reward from these naturalistic feedback expressions.

\Paragraph{Facial Emotion Recognition.} We estimate the facial emotion expressions in each frame with the facial emotion classifier provided by the \textsc{fer} package~\cite{Arriaga.20.10.2017,Arriaga.2020}, employing the MTCNN face detector~\cite{Zhang.2016} which has been trained on a large-scale facial expression dataset~\cite{Goodfellow.2013}. 
For each frame, \textsc{fer} predicts a discrete probability distribution characterized by the probability vector $\mathcal{P}(E)$, which describes how likely the user's facial expression corresponds to each of the seven universal facial emotions established by \cite{Ekman.1987}. The emotions are defined in the set $E$, which we classify as: positive (`happy'), neutral (`neutral', `surprise'), and negative (`angry', `disgust', `fear', `sad') feedback. 

%as summarized in Table~\ref{tab:emotion_scaling}. %: 'angry', 'disgust', 'fear', 'happy', 'sad', 'surprise', and 'neutral'. %, given its observation (i.e., the photo or video frame): $\mathcal{P}(E)$ %or $\mathcal{P}(E | \text{frame})$, which holds the individual probabilities for each emotional expression $e \in E$, such as $p\,(e = \text{'angry'} | \text{frame})$.
% \begin{figure}[ht]
% 	\centering
% 	\includegraphics[width=\columnwidth]{figures/example_prob_dist.pdf}
% 	\caption{Example emotion probability distribution output by \textsc{fer}.}
% 	\label{ferex}
% \end{figure}
%However, what we are actually interested in is \emph{how satisfied is a user $u_i$ with the drone's chosen action $\ac$}. Thus, we somehow need to infer the user's satisfaction from these facial expression estimates.%, i.e., the probability that s/he was satisfied given the facial emotion probability distribution $\mathcal{P}(E)$:
% \begin{equation}
%     p\,( u_i = \text{satisfied} | \mathcal{P}(E), \ac)
% \end{equation}
% other terms, 

\Paragraph{Inferring the Agent's Reward.} 
Essentially, the desired \emph{reward signal} $r$ corresponds to the user's \emph{level of satisfaction}, a latent variable that cannot be directly accessed but is inferred from the output of facial emotion recognition. %: How (dis)satisfied was s/he with the drone's performed action?
%However, this represents a \emph{latent variable}, since we cannot directly access the user's mental satisfaction state. Rather, we need to estimate this hidden state from surrogate observations, notably the user's facial expressions. % during providing feedback. % (note that the human user also will explicitly seek to convey their satisfaction by providing feedback to the drone). 
%Therefore, we base on Ekman's established, universal facial emotion classification system~\cite{Ekman.1987,Ekman.}, comprising a set of seven emotions, $E$: the six cross-cultural emotional base expressions ('angry', 'disgust', 'fear', 'happy', 'sad', 'surprise') as well as a 'neutral' expression. 

The RL problem requires a scalar reward value $r$. Thus, we define a mapping of the emotion probability vector $\mathcal{P}(E)$ to $r$ that conveys the idea of satisfaction or dissatisfaction of the user. $r$ corresponds to the \emph{emotional valence} score, representing the extent to which an emotion is positive or negative~\cite{Citron.2014,Kossaifi.2021}. In \emph{affective computing} terms, this realizes a mapping from Ekman's basic discrete model to the pleasure-displeasure scale of the continuous valence-arousal-dominance model~\cite{Mehrabian.1974}.%, similarly to the approach proposed in~\cite{BuechelSven.2016}. 

In detail, the scalar reward $r$ is calculated as the dot product between the emotion probability vector $\mathcal{P}(E)$, and its corresponding scaling factors vector $\mathbf{s}$:

\begin{equation}
    r = \mathcal{P}(E) \cdot \mathbf{s},
    \label{eq:scorer1}
\end{equation}

where $\mathbf{s}$ is configured as $[$+3 (happy), 0 (neutral),  +1 (surprise), -3 (angry), -2 (disgust), -2 (fear), -3 (sad)$]^T$. Thus, \emph{positive} emotions will increase, \emph{negative} emotions will decrease the resulting score $r$. Note that alternative calculations are possible, e.g., just selecting the most probable emotion, yet this yielded worse results in our first experiments.

\Paragraph{Implementing Agent Learning.} %Since we are dealing with an \emph{action selection task with evaluative feedback}, 
We  formulate our learning problem as a %contextual
\emph{multi-armed bandit (MAB) problem}~\cite{Sutton.2018}:
Each command $\co \in \Co$ represents a context,\footnote{Note that the corresponding context is directly observable and hence does not have to be inferred, thus our problem corresponds to a classical MAB setting involving multiple bandits rather than a contextual/associative bandit.} 
thus instantiates an MAB with $k$ arms. Each arm represents one particular action $\ac \in \Ac$, whereby we will denote these actions by indexing their enumeration in vector $[a_1, \ldots, a_k]^T$. 
%, each action $\ac \in \Ac$ corresponds to a bandit arm
The user's satisfaction with a chosen action $a$ is expressed by the reward $r$. The drone selects the action that is expected to maximize its collected reward $\sum_T r$ over time $T$, in other words, the action with the highest expected (reward) value.

More precisely, the MAB estimates the actions' values as
$Q_t(\ac)$ by consistently updating the mean rewards received for each of its actions up to its current trial step $t$. 
\begin{figure}[t]
	\centering
	\includegraphics[width=\columnwidth]{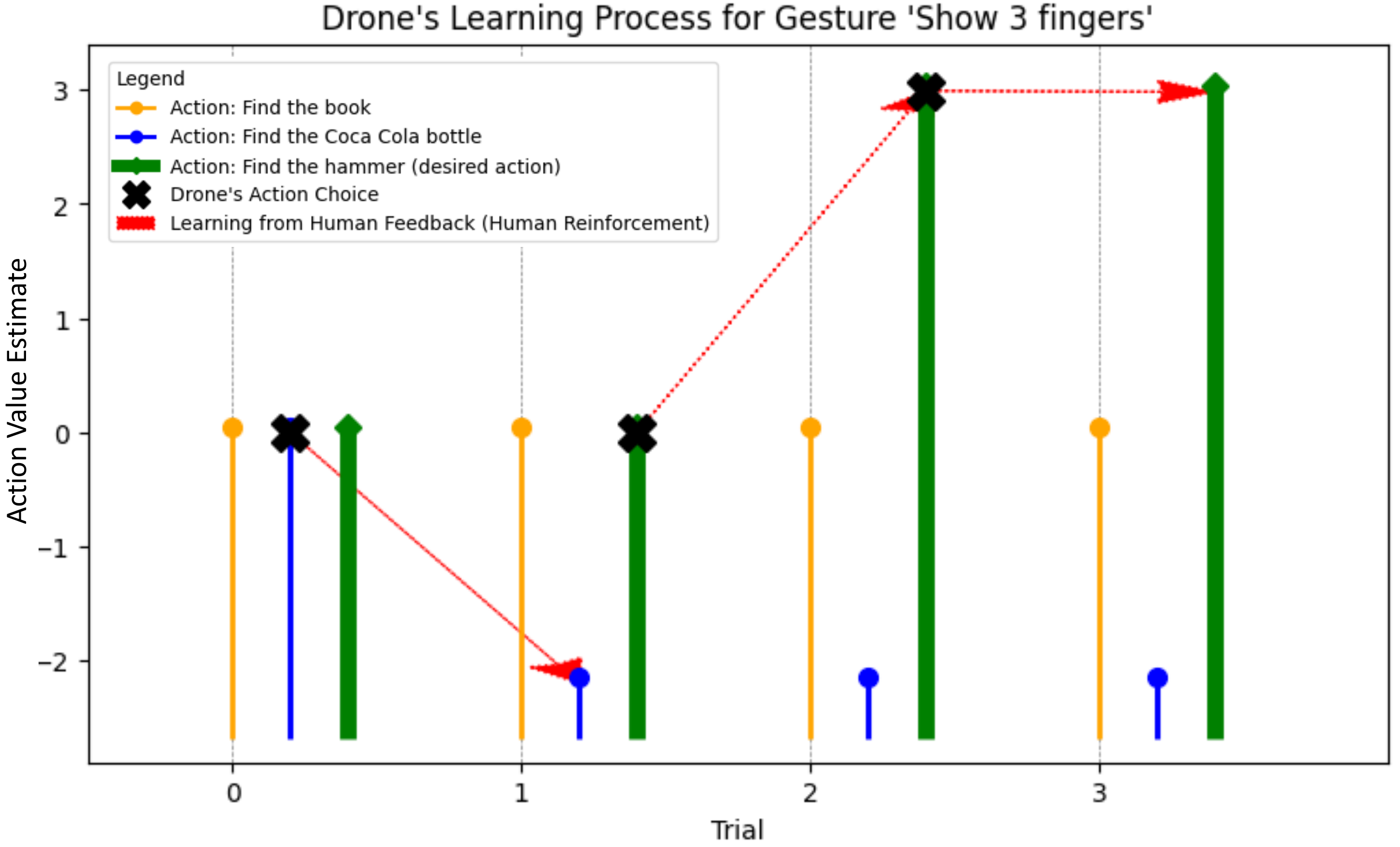}
	%figures/learning_analysis_webplot2.png}
	\caption{Drone agent's learning evolving over time based on the user's interactive feedback, ideal sample use case (taken from our study with three possible action choices, user ID 13): thick line indicates the desired action.}
	\label{fig:learninganalysis}
\end{figure}
%\Paragraph{Examining the Interactive Learning Process.}
Fig.~\ref{fig:learninganalysis} depicts an example from our user study for inspecting the drone's learning. It is shown how the drone's action value estimates $Q_t(\ac)$ evolve with repeated human reinforcement through emotional rewards $r$. At each trial step (x-axis), the drone selects the action (bandit arm) with the current maximum action value $Q_t(\ac)$ (y-axis), or randomly samples from the maximum ones in the case of ties.\footnote{We initially implemented classical $\epsilon$-greedy action selection: In $1-\epsilon$ \% of the trials the action with the max. $Q(\ac)$ is chosen (\emph{exploitation}), in $\epsilon$ \% a random action is selected (\emph{exploration}) to refine $Q(\ac)$ estimates. After initial tests, we chose to disable  \emph{exploration} and realize deterministic action selection by setting $\epsilon = 0$, since in our setting learning will be actively ``steered'' by the user.} 
At trial step $t=0$ for this MAB, the drone thus selects an action randomly out of three actions.  %$[a_1, a_2, a_3]^T$. %\andrea{($a_2$)}. 
This action is not the desired one and the negative user feedback leads to a decrease of this action's value estimate.  %\andrea{$Q_1(\ac_2)$}. 
At trial step $t=1$, the drone selects the correct action  %\andrea{($a_3$) }
out of the two  actions with maximum value, leading to positive user feedback. From this step on, the drone has learned the correct command-action mapping.

As can be seen in this example, the MAB is initialized with action values $Q_0(\ac)$ of \emph{neutral valence} (0). We also experimented with an \emph{optimistic initial values} approach~\cite{Sutton.2018} where all action values are initially set to $+5$. %, by initializing the action values with overly optimistic action values (all set to $+5$). 
This forces the MAB to first try out each action at least once to collect the user's real feedback, even if the user's desired action is selected early-on, as in Fig.~\ref{fig:learninganalysis}. To reduce the necessary number of interactions, we thus decided to use values of neutral valence.

\section{Experiments}

We conduct an experimental user study in the form of a custom Web app\footnote{\url{https://apps.tk.jku.at/he4dc/}} emulating the drone in the interactive learning loop, see Fig.~\ref{fig:webappexp}.

\Paragraph{Experiment Setup.} The study considers three command finger gestures and three actions (i.e., $k=3$): finding a hammer, a book, or a bottle. The drone is required to learn each of the command-action mappings in one teaching session. The study participants 
first specify their desired command-action mapping %for the session 
(\emph{ground truth} mapping), then the following teaching loop is performed: (1) the user selects a gesture command; (2) the (emulated) drone chooses the action with the maximum action value (or, if multiple actions with maximum action value exist, chooses one of those actions randomly) and a video is played emulating the drone's flight towards the anticipated object to be found; (3) the user provides real facial emotional feedback recorded by a Web cam over a time period of 5~seconds and labels his/her emotional response as positive or negative (\emph{ground truth feedback}). 

With this setting, we abstract from the embodied hardware agent to limit the learning confounders and focus on the evaluation of the usefulness of the facial emotional feedback per-se. Yet, we performed initial experiments with our hardware drone (Parrot Anafi) to make sure that video processing (frame selection and \textsc{FER}-based emotion classification) and taking a video with the on-board camera of the drone is of comparable video quality. This is important as we plan to %follow up on this study
extend the investigation by a lab-experiment with a real drone, where we will study the influence of practical factors such as light conditions and camera angle.

%perform the following sequence until they are convinced that the drone agent has correctly learnt their desired command-action mapping: The participant selects one of the available gestures, then will be shown a video of the drone's taken action, and consequently will be able to submit their emotional feedback using the webcam of their device. 
% \begin{figure}[h]
% 	\centering
% 	\includegraphics[width=\columnwidth]{figures/webapp.png}
% 	\caption{The interactive machine teaching loop implemented in our web app-based drone training experiment.}
% 	\label{fig:exp03}
% \end{figure}
%Hence, this setting allows us to involve a larger number of human participants in our virtual experiment than would have been possible with the physical experiment to be held in our lab later on. Moreover, 

%\todo{karin: The following is a repetition of what we argued before, maybe deleting? ... By abstracting away many of the peculiar challenges of involving a hardware-embodied agent, this setting enables a suitable proof-of-concept study for assessing the overall feasibility of our approach as well as guiding the parameter settings for the later experiments involving the physically embodied drone agent.}

\begin{figure}[t]
     \begin{subfigure}[t]{0.45\columnwidth}
         \centering
	\includegraphics[width=\textwidth]{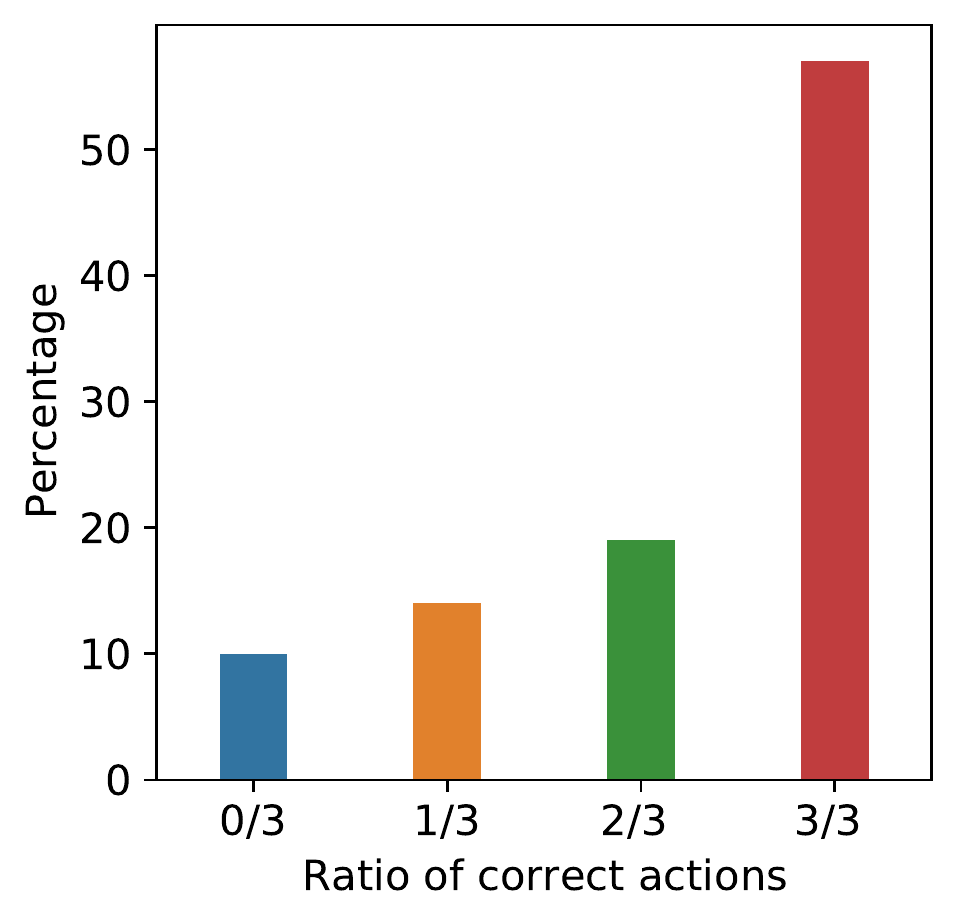}
	\caption{Drone teaching success rates. %\todo{Manuela: enlarge text; align x-axis to x-axis of fig. (b) -- best also top line as well, capital first letters; change times to percentage (i.e., e.g. 3/3: about 57\% ... make sure that it sums up to 100 overall.} %\andrea{Would also specify the y-axis on the \%-range, would make the plot higher (aspect ratio, given via fig size params), the bar widths thinner, would suggest writing the x-axis labels as ``text'' in 90° angle over/next to the bars (labels: ``2/3 actions correct'', ``3/3 actions correct'' etc. $\to$ allows to increase font size!}
	}
	\label{fig:exp03}
	\end{subfigure}
	\hfill
	\begin{subfigure}[t]{0.48\columnwidth}
	   \centering
        \includegraphics[width=\textwidth]{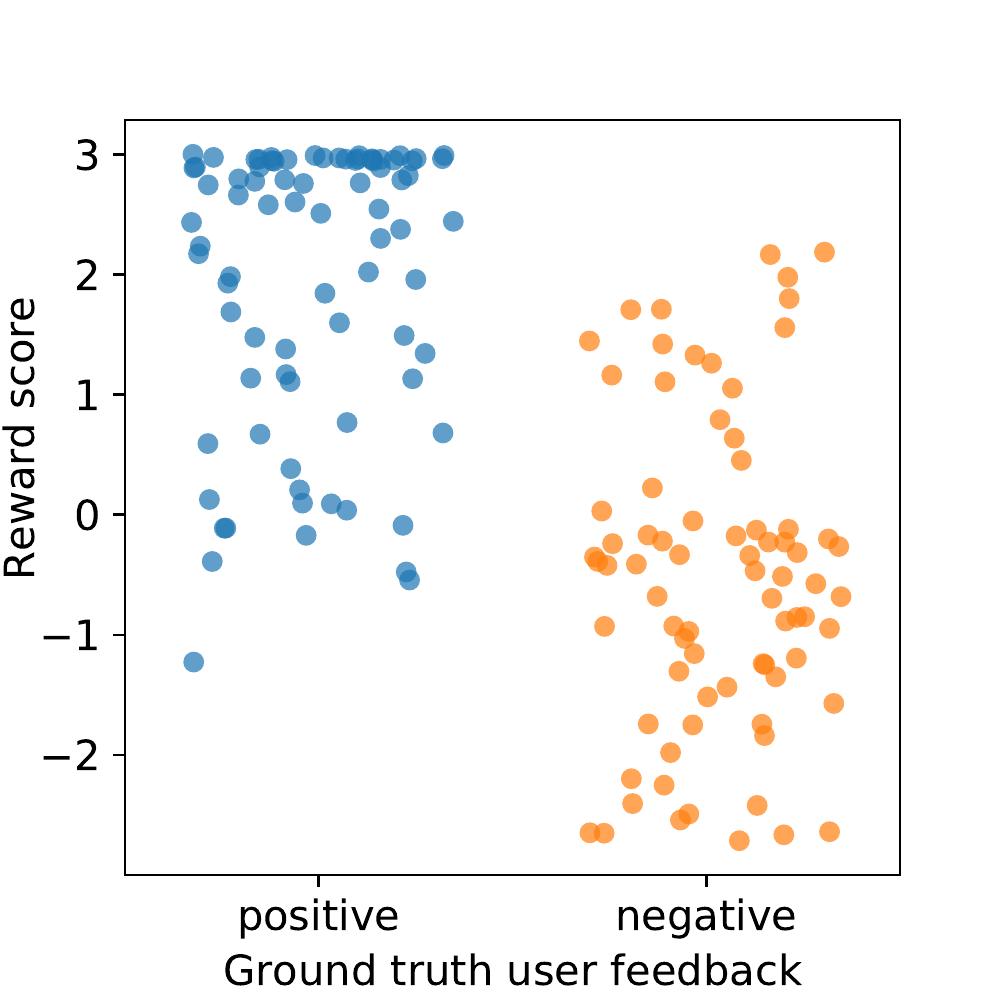}
	\caption{Reward scores %of each feedback round 
	wrt. ground-truth feedback.}
	\label{fig:defb-groundtruth}
	\end{subfigure}
	\caption{Empirical results on drone teaching success.}
\end{figure}

\Paragraph{Video Frame Rate and Reward Calculation.} 
The frame rate of the video stream is 25 frames per second (FPS), which is down-sampled for emotion classification 
by only retaining every $j$-th frame. We set $j=12$, which makes real-time emotion classification feasible that takes about 32-42~seconds per face. 
%\footnote{
%\subsubsection{Temporal Fusion of Emotion Estimates.} 
%Our drone obtains a 
%video stream, i.e., sequence of frames, at a frame rate of 25 FPS. Since on our server's hardware, emotion classification on a frame containing a face takes around 32-42 seconds, we need to \emph{downsample} our obtained stream of frames to allow for real-time processing. Thus, we only select every $j$'th frame for downstream processing, according to a configurable step-size parameter $j$ (given our server's processing times, we set $j= 12$). %, thus reduce the number of frames to process from our downsampled set $\mathcal{I}_{F}$ to~$F = \frac{N}{j}$. %We also experimented with a \emph{smoothing} step on our emotion estimates after this \emph{filtering} operation  to mitigate the effect of outliers (i.e., frames which have a substantially different classification than their neighbouring frames), by computing a moving average over sliding windows, using overlapping windows (we use a sliding window size of \todo{$w = ?$} preceding frames and a window overlap of \todo{$w_o = ?$} frames from $\mathcal{I}_{F}$). However, 
%Since our real-world experiments indicated that emotional responses in short time windows were sufficiently stable, we have eventually discarded our initially attempted smoothing steps for performance reasons. 
%Thus, we finally compute 
We then compute the \emph{average emotion probability vector} $\overline{P(E)}$ by computing the mean of the derived sequence of $\mathcal{P}(E)$s obtained on the frames of the down-sampled set. The overall emotional reward $r$ is calculated by applying Eq.~\ref{eq:scorer1} to $\overline{P(E)}$.

\Paragraph{Study Participants and Sessions.} 16 %individual 
volunteers are included in the discussion, all with an academic background, among them 56 \% female and 44 \% male, who performed in total 21 drone teaching sessions with a total of 157 feedback rounds.\footnote{We did not include sessions not completed by the respective user in order not to bias the data. %, meaning that 
Thus, only sessions are included where each of the three commands has been selected at least once.}

\begin{figure*}[!t]
	\centering
	\includegraphics[width=\textwidth]{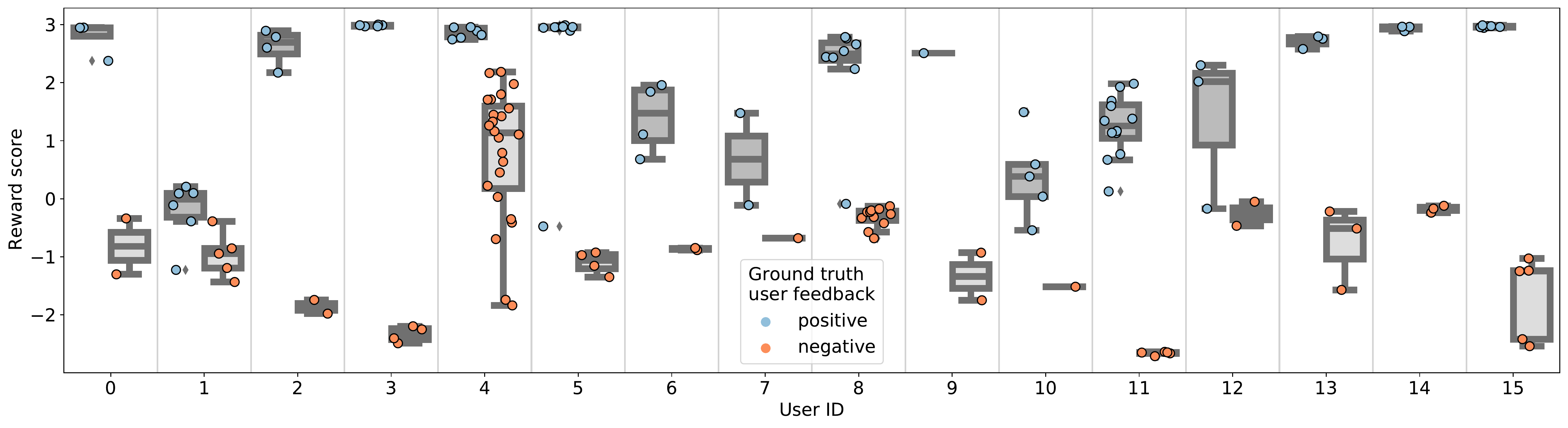}
	\caption{Comparison of the reward score distributions of positively and negatively intended feedback (ground truth feedback) for each study participant. The boxes in the boxplot show the median and the 25 \% and 75 \% quantiles.}
	\label{fig:defb-groundtruth-boxplots}
\end{figure*}

%------------------------
\section{Analysis and Discussion}

We study whether our approach based on emotional feedback overall leads to successful drone learning and whether the scores derived by our \textsc{fer}-based RL algorithm correlate with the users' intention (ground truth feedback). %Then, 
We dissect this result further by analysing the capabilities of \textsc{fer} per-se. Finally, we quantify the quality of learning, i.e., assess the magnitude of how well desired and undesired actions can be distinguished in the drone's learned action values.

\Paragraph{(1) Did users succeed in teaching the drone their desired command-action mapping?} -- %, solely based on emotional feedback?}
%As an encouraging finding, with our devised learning approach 
The majority of users successfully taught the drone agent %their desired command-action values, 
based on a few trials only (see Fig.~\ref{fig:exp03}): %Of all sessions where all three possible commands were trained, 
$57$\% of the training sessions succeeded to correctly teach all three command-action mappings and $19$\% succeeded in correctly teaching two out of three; only less than $10$\% of the sessions yielded in no success at all. These first results indicate the overall feasibility of our approach using emotional response as a reward signal.  
%\todo{Success Rate \ldots}
% \begin{figure}[h]
% 	\centering
% 	\includegraphics[width=0.5\columnwidth]{figures/defb-groundtruth.pdf}
% 	\caption{Distributions of reward scores wrt. ground truth}
% 	\label{fig:defb-groundtruth}
% \end{figure}

% \begin{figure*}[t]
% 	\centering
% 	\includegraphics[width=\textwidth]{figures/doefb_bp.pdf}
% 	\caption{Comparison of the reward score statistics of positively and negatively intended feedback (ground truth label) for each study participant. The boxes in the boxplot show the median and the 25\% and 75\% quantiles.}
% 	\label{fig:defb-groundtruth-boxplots}
% \end{figure*}
\Paragraph{(2) Are the reward scores correlated with the ground truth feedback?} -- 
%One crucial question for our approach is whether emotional responses -- and our reward scores derived therefrom -- are sufficiently similar between different subjects in order to serve as a generic learning signal.
We can analyze whether the reward scores based on 
%, generically computed from the users' recorded video feedback based on 
Eq.~\ref{eq:scorer1} correlate with the ground truth feedback labels provided by the users (positive or negative rating of the drone's chosen action). %Based on our collected ground truth labels, w
%indicating their sought-after feedback type. 
Fig.~\ref{fig:defb-groundtruth} shows the reward scores wrt. their corresponding ground truth label across all trials and participants. %, aggregated over all study participants. 
The distributions of positively and negatively intended reward scores indeed differ, as visually observable, which is confirmed by a two-sample Kolmogorov-Smirnov test that rejects the null hypothesis $H_0$ with $p < 0.001$ ($H_0$: The positively and negatively intended reward scores come from the same distribution). Yet, we also see that the reward scores of a considerable amount of negatively labeled scores have positive values. %\todo{Just to be sure: these are only data from our 16 participants, right?} \remark{Yes. I would omit the previous sentence, since we never really claimed that negatively intended scores should have negative values -- our algo. works on the \emph{relative} distances between the scores...}

%, which is statistically \emph{highly significant}\footnote{A Two-sample Kolmogorov-Smirnov Test rejects the null hypothesis $H_0$ that the positive and the negative reward scores come from the same distribution with $p < .001$}.
% \begin{wrapfigure}{r}{4cm}
% 	\centering
% 	\includegraphics[width=0.5\columnwidth]{figures/defb-groundtruth.pdf}
% 	\caption{Distribution of reward scores wrt. ground truth}
% 	\label{fig:defb-groundtruth}
% \end{wrapfigure}
To dissect this phenomenon, we analyze the score distributions of each individual participant,
%The picture becomes even clearer when we analyze these score distributions per subject
as shown in Fig.~\ref{fig:defb-groundtruth-boxplots}. %, revealing strong individual differences between participants. 
We generally observe a visible difference between the distributions of the positive and negative scores, indicating that our generic reward score computation (Eq.~\ref{eq:scorer1}) produces a suited learning signal. Yet, we find that both positive and negative score medians differ widely between the different participants. 
%On the per-subject level, the distributions between positive and negative reward scores are typically highly separated, indicating that our generic reward score computation (Eq.~\ref{eq:scorer1}) produces a suited learning signal. However, we also note strong individual differences between users in the numeric range of their feedback distributions, and how well their feedback score distributions are separated.
For example, whereas the positive and negative feedback distributions of the participant with user ID 1 appear rather weakly separated, users with IDs 2--3 and 13--15 show extremely well separated positive and negative feedback distributions. %To better understand why these differences between users exist, we evaluate the performance of emotion recognition in the following. 
\begin{figure}[t]
        \centering
	\includegraphics[width=0.9\columnwidth]{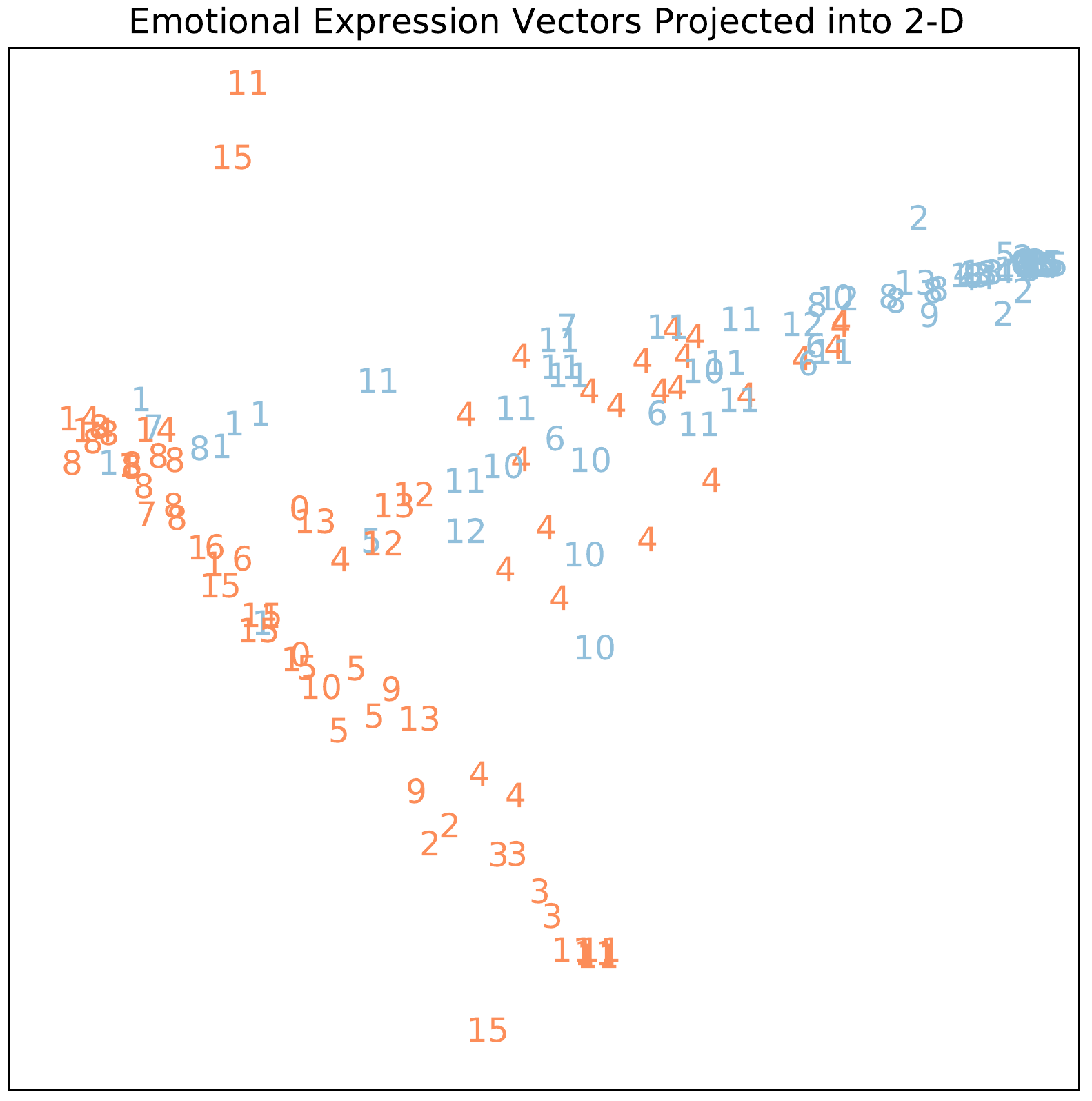}
	\caption{7-d emotional response vectors projected into 2-d for all user IDs and colored wrt. ground truth feedback (blue: positive, orange: negative).}
	\label{fig:mds01}
\end{figure}

\begin{figure}[h]
        \centering
	\includegraphics[width=0.9\columnwidth]{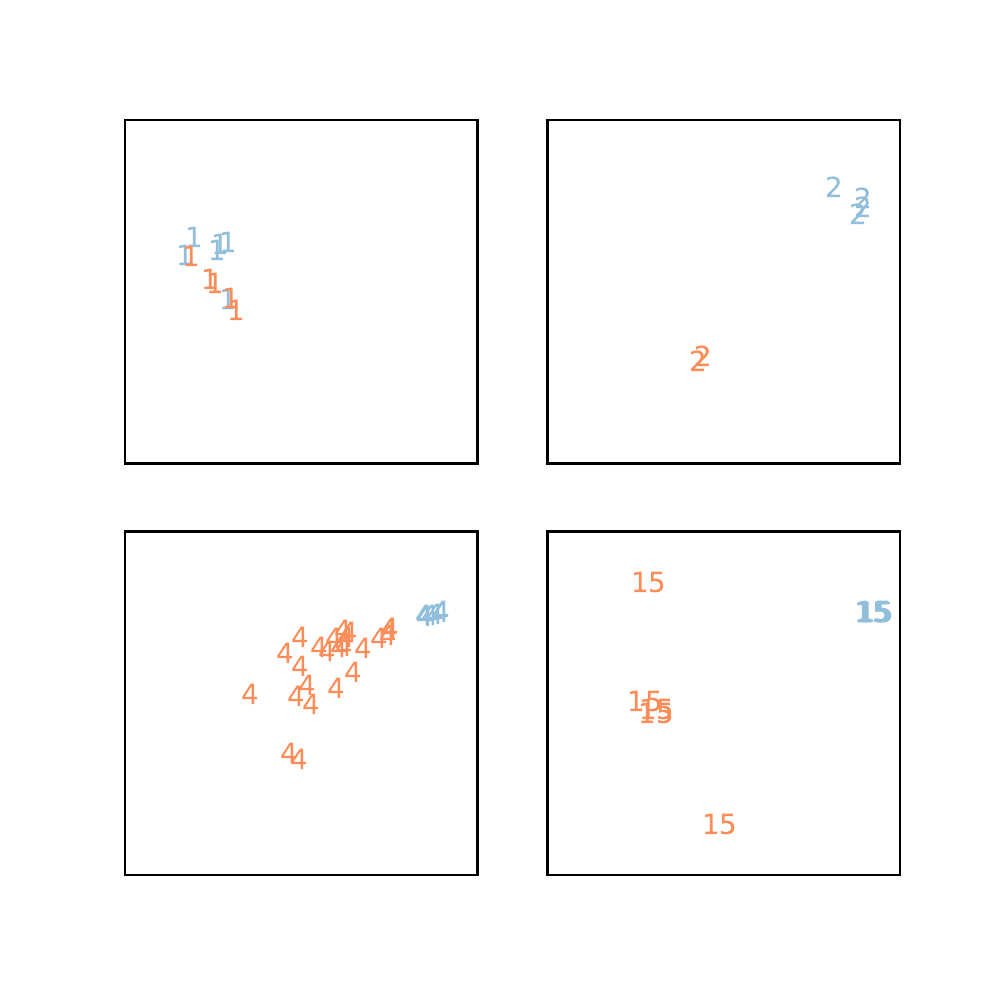}
	\caption{Facetted view of Fig.~\ref{fig:mds01}, outlining selected participants' 2-d projection of the 7-d emotional response vector.}
	\label{fig:mds02}
\end{figure}

% \begin{figure}[h]
% 	\centering
% 	\includegraphics[width=0.7\columnwidth]{figures/mds02.pdf}
% 	\caption{Facetted view of Fig.~\ref{fig:mds01}, outlining selected participants' characteristics.}
% 	\label{fig:mds02}
% \end{figure}
\subsubsection{(3) Are the facial emotion classifications correlated with the ground truth?} -- 
We now evaluate the initial emotional responses before applying the reward function. 
%Next, we take a step back from our custom defined reward computation function to investigate the initially provided emotional responses. %We performed \emph{Multi-Dimensional Scaling}~(MDS)~\cite{Kruskal.2009} on the input for our reward signal computation, i.e., the resulting seven-dimensional aggregated emotional response vectors $\overline{P(E)}$. 
Fig.~\ref{fig:mds01} and Fig.~\ref{fig:mds02} visualize the 2-dimensional projections of the 7-dimensional average emotional probability vectors $\overline{P(E)}$, %\todo{consistent way of talking about P(E); is it a vector? is it a distribution?; \andrea{it is \emph{both} -- in ML, it's common to write a categorical distribution as a probability vector, since it should sum to 1 which then can easily be checked (as the sum of the vector's elements), \url{https://en.wikipedia.org/wiki/Probability_vector}} when do we use overline and when not?} \andrea{overline = mean of our downsampled P(E) vectors = averaged emotion distribution of our entire (downsampled) feedback video, conventional x-bar notation, https://en.wikipedia.org/wiki/Mean -- though yes, it looks a bit strange with the upper-case distribution under it..., but the problem is that the convention is to use upper-case $P$ for the entire probability distribution and lower-case $p$ for an individual probability, so don't see how to resolve that w. sticking to conventional notation}
labeled according to the ground truth feedback labels (positive/negative). 
%project our participants' originally provided emotional response $\overline{P(E)}$ vectors with 
The projection is achieved with \emph{multi-dimensional scaling}~(MDS)~\cite{Kruskal.2009}, a method that aims to preserve the %ensures that
distances between the data points in the lower-dimensional projection. %are retained.
%into a 2-D space, for outlining the distances between the different emotion vectors. Also on the level of the seven-dimensional 
It can be observed that positive and negative emotional responses tend to cluster across different subjects. However, we also note an overlapping area where a clear separation is not possible, mainly caused by the individual effects of a specific user (User ID 4). 

%\karin{COMMENT: I would not argue again with the boxplots as we are here saying we want to discuss the raw emotional responses}
%this grouping is not clearly separable on this aggregate, across-subject level: For example, the negative scores of user no. 4 (lying in the center region) are closer to other participants' positive scores, as also evidenced by comparing this participant's box plot in Fig.~\ref{fig:defb-groundtruth-boxplots}, revealing that this person's negative scores are the highest ones of all participants. 

To provide a quantitative assessment of the general correlation between the emotion vectors and the ground truth labels, we fit a \emph{logistic regression} model onto the original 7-dim. average emotion vectors $\overline{P(E)}$ to predict the ground truth feedback labels (positive/negative), which results in a prediction error of $20$\%,  which can be interpreted as a quantification of the overlap between the positive and negative feedback across different users, likely corresponding to the interspersed area of positively and negatively labeled emotional responses in Fig.~\ref{fig:mds01}. %for predicting the training dataset's labels 
%To explain the root cause of this phenomenon, 
Fig.~\ref{fig:mds02} dissects this 2-dimensional projection for four selected individual participants, revealing typically well-separated ``clusters'' of positive and negative emotional responses, but also outlining a user with emotional characteristics that are difficult to distinguish (User ID~1). 
%\karin{One can observe that the shape of the clusters differs both in inter-cluster and intra-cluster distances}. 
%(which, in a sense, ``quantifies'' the overlap between the positive and negative feedback across different users). Conversely, the subject-level analyses depicted in Fig.~\ref{fig:defb-groundtruth-boxplots} and 
%Fig.~\ref{fig:mds02} suggest different ``decision boundaries'' for different users, 
These results indicate the need for learning a \emph{user-specific} feedback inference model, and thus, confirm the fit of our RL-based approach to the emotional feedback case. Further, our MAB-approach for rewarding allows to consider \emph{relative distances} of each individual user's feedback scores, thus mitigating the problem of different individual inter-cluster distances of emotional vectors.

\subsubsection{(4) How well have users been able to teach the drone?} 
\begin{figure}[t]%{0.4\textwidth}
\centering
\includegraphics[width=\columnwidth]{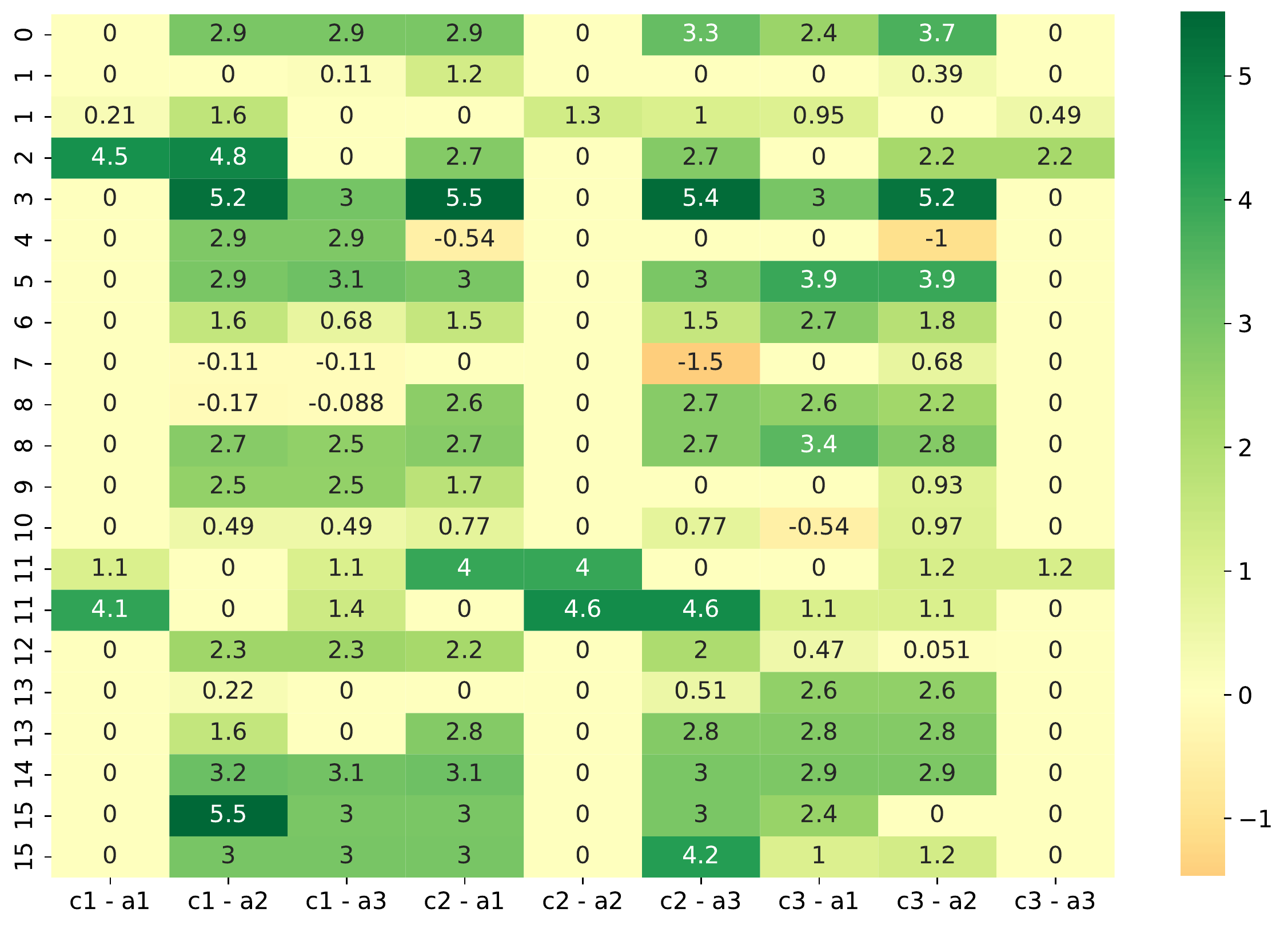}
\caption{
Difference of action values to the desired action $a_d$, $d(a_i) = Q_T(a_d) - Q_T(a_i)$, across all command-action combinations (x-axis) of all sessions (y-axis, labeled by participant ID -- note that one participant may have conducted multiple sessions): yellow cells correspond to $d(a_i) = 0$ (as given for $a_{i=d}$); 
%\andrea{indicating} the desired action \andrea{$a_{i=d}$}; 
the greener the cell, the larger $d(a_i)$. %, i.e., the greater the difference between \andrea{this undesired action the desired one}. 
Orange color indicates wrongly learned actions. %Note that one participant may have conducted multiple sessions.
%\karin{Difference $d(a)$ between each action's reward score and the desired action's reward score, across all command-action combinations (x-axis) of all session identified by participant ID (y-axis):} yellow cells correspond to $d(a) = 0$\andrea{, thus indicate the desired action}; the greener the cell, the larger is $d(a)$, i.e., the difference of the reward score of the (undesired) action and the desired action. Orange/red colors indicate wrongly learned actions. %\remark{I'd still think it might make sense to put it as 3rd sub-plot in the row of Fig. 5 -- there's quite some room for further optimizing all those figs. (e.g., cells widths of the heatmap could be made slimmer), and then one has all plots nicely together for comparing participant-effects across plots.}
%\todo{aspect ratio + font should be improved to better fit in - make higher, decrease cell width, increase font size, esp. for x- and y-axis labels?}
}
\label{fig:heatmap02}
\end{figure}

-- The MAB-based approach allows the drone to learn a command-action mapping even when the difference between the average rewards for desired and non-desired actions is small. In our last analysis, we thus inspect the drone agent's \emph{final learning outcome} at the last time step $T$ of each teaching session, by quantifying 
the difference between the participant's \emph{desired} and \emph{undesired} actions, with actions referred to by their position index in vector $[a_1, \ldots, a_k]^T$. The final action values $Q_T(\ac)$ represent the drone's ``view'', where  %In terms of Fig.~\ref{fig:learninganalysis}, the difference between the bars' heights in each participant's last trial step for each command. Ideally, 
the \emph{desired} action $a_d$ should ideally yield the maximum action value estimate $Q_T(\ac_d)$. 
%, whereas undesired actions \andrea{$a_{i \neq d}$} ideally have low values. %To quantify this difference
%for an action $a$
Fig.~\ref{fig:heatmap02} visualizes the difference
$d(a_i) = Q_T(\ac_d) - Q_T(\ac_i)\quad \forall\, 1 \le i \le k$ % and $r()$ is the reward function \andrea{from Eq.~\ref{eq:scorer1}.
across all final command-action mappings in all sessions.
%Note that one participant may have conducted multiple teaching sessions. 
%\karin{clarify here the multiple user IDs; User 2 not successful comment of reviewer 3} %\todo{speak about difference, not distance as we also have negative distances}

%We also specifically analyze these ``distances'' in the drone's finally learned action values: 

It can be observed that some sessions show a strong pattern, i.e., $d(a_i)$ is large for all undesired actions $a_{i \neq d}$, meaning that 
%quantifies the distance of  the user's intended action's value to the undesired actions' values (i.e., viewed in terms of Fig.~\ref{fig:learninganalysis}, plots the difference between the bars' heights in the last trial step), across teaching sessions 
%the user's 
teaching was efficient (which corresponds to well-separated feedback scores as given in Fig.~\ref{fig:defb-groundtruth-boxplots}). 
Conversely, negative values of $d(a_i)$ indicate that a non-desired action has received a higher action value than the desired action, indicating a wrongly learned mapping. Based on this color-coding, we can easily identify unsuccessful teaching, %but also those study 
and participants (e.g., user ID 2 and 15) who have been most effective in  successfully instructing the drone as well as participants who have given rather weak feedback signals (e.g., user ID 1). This again reveals considerable individual differences among participants.
%Difference between ground truth and reward score. If the value is 0 (yellow color) the drone has learned right action. the higher the value the more positive was the false emotional feedback to this action - the lower the value the more negative was the emotional feedback. 

%\karin{karin stopped here}

% \begin{figure}[h]
% 	\centering
% 	\includegraphics[width=\columnwidth]{figures/dbgr_wm_heatmap.pdf}
% 	\caption{The heatmap shows the difference between the ground truth's action value and the undesired actions' values across all possible command actions possibilities. The greener, the more pronounced was the feedback difference on \emph{desired} vs. \emph{undesired} action selections. Red indicates wrongly learned actions. Yellow marks the ground truth action.}
% 	\label{fig:heatmap02}
% \end{figure}

%\todo{maybe also collect user's feedback $\to$ questionnaire on Likert scale?}
%\clearpage

\section{Conclusion}

We contributed insights on developing artificial agents that learn based on human emotional feedback, studied along the use case human-drone interaction. We conducted a user study with 16 participants. 
%in which machine and human interact based on the humans' natural communication channels. 
Our initial empirical findings confirm that humans' facial emotion expressions indeed can be exploited as reward signals in interactive human-in-the-loop reinforcement learning. %We were specifically surprised to find that our generic mapping function for converting the emotion distributions into a numeric reward score generally worked ``out-of-the-box'' for most of our participants,
For the majority of the study participants, teaching the drone worked successfully, supported %by our approach and 
in particular by our reward function derived from emotional feedback. The learning process accounts for individual differences in expressing emotions, without necessitating any further personalization to individual users. %' characteristics. 
%\manuela{Future Work}
%\manuela{In this study, we used a vector of emotion for the RL settings. In order to achieve more meaningful results, a switch to multi-reward reinforcement learning is planned.}\karin{let's formulate the change from a single value to a multi-reward RL approach as an option for future work (indeed interesting though for our future work)}

%\karin{Another interesting investigation (beyond the scope of the work): compare emotional feedback with thumbs-up/thumbs-down signalling (with noise)}

This is a first study on the feasibility of emotion-based reinforcement learning for flying robots. Many practical aspects are subject to future studies, including the limitations of emotional feedback wrt. complex tasks, the inaccuracies introduced by noise and low-quality video feed, and that observing human emotion may be a threat to privacy. 

\section*{Acknowledgements}
We sincerely wish to thank all our study participants. %[\emph{anonymized}] was funded by grant [\emph{anonymized}]. 
A.~Salfinger was funded by the Austrian Science Fund
(FWF) under grant FWF T961-N31. 
%from [\emph{anonymized}].
%\clearpage

%\newpage

%\bibliographystyle{ACM-Reference-Format}
\bibliography{lit}

\end{document}